\documentclass[journal]{IEEEtran}
\usepackage{epsfig,amssymb}
\ifCLASSINFOpdf
\else
\fi
\begin{document}
\font\scten=msbm10 \font\sctwelve=msbm10 scaled 1200
%

\title{\Huge Multivariate Interpolation Formula over Finite Fields and Its Applications in Coding Theory}

\author{Yaotsu Chang, Chong-Dao~Lee, and Keqin Feng
\thanks{This work was supported by National Science Council, R.O.C., under Grants NSC99-2115-M-214-002-MY3 and NSC99-2221-E-214-051-MY3. The third author is supported by the grant of the  NSFC no.10990011 and the Tsinghua National Lab. of
Information Science and Technology.}
\thanks{Y. Chang is with the Department
of Applied Mathematics,I-Shou University, Dashu Township, Kaohsiung Country 840, Taiwan, R.O.C. (e-mail: ytchang@isu.edu.tw).}
\thanks{C. D. Lee is with the Department
of Communication Engineering, I-Shou University, Dashu Township, Kaohsiung Country 840, Taiwan, R.O.C. (e-mail: chongdao@isu.edu.tw).}
\thanks{K. Feng is with the Department of Mathematical Sciences, Tsinghua University, Beijing 100084 China (email: kfeng@math.tsinghua.edu.cn)}
}

\markboth{}%
{Chang \MakeLowercase{\textit{et al.}}: Bare Demo of IEEEtran.cls
for Journals}

\maketitle

\begin{abstract}
A multivariate interpolation formula (MVIF) over finite fields is
presented by using the proposed Kronecker delta function. The MVIF
can be applied to yield polynomial relations over the base field
among homogeneous symmetric rational functions. Besides the
property that all the coefficients are coming from the base field,
there is also a significant one on the degrees of the obtained
polynomial; namely, the degree of each term satisfies certain
condition. Next, for any cyclic codes the unknown syndrome
representation can also be provided by the proposed MVIF and also
has the same properties. By applying the unknown syndrome
representation and the Berlekamp-Massey algorithm, one-step
decoding algorithms can be developed to determine the error
locator polynomials for arbitrary cyclic codes.
\end{abstract}


\IEEEpeerreviewmaketitle

\section{Introduction}

\subsection{Earlier Work}
\label{History}

Cyclic codes, proposed by Prange in 1957
\cite{IEEEhowto:Prange57}, are important and practical error
correcting codes. They are widely used today, including the
Bose-Chaudhuri-Hocquenghem (BCH) codes and Reed-Solomon (RS)
codes. Many decoders have been developed since then, including the
famous Berlekamp-Massey (BM) algorithm which was designed
specifically for the BCH codes or RS codes at first. Actually,
other cyclic codes can also be decoded by the well developed BM
algorithm, provided that there are enough consecutive known
syndromes; namely, $2t$ consecutive syndromes are needed to
correct a corrupted word with at most $t$ errors. Unfortunately,
for any cyclic codes other than the BCH/RS codes, the number of
consecutive known syndromes is less than $2t$. To obtain the
unknown syndromes, Feng and Tzeng \cite{IEEEhowto:Feng94} proposed
a matrix method which yields expressions of unknown syndromes in
terms of known syndromes. With the extra syndromes, some BCH codes
can be decoded up to their true error correcting capacity rather
than the smaller designed one. Later, He et al.
\cite{IEEEhowto:He47} developed a modified version of Feng's
method, and used it to determine the needed unknown syndrome and
then to decode the binary quadratic residue (QR) code of length
47. By applying the Feng-He's matrix method accompanied by the BM
algorithm, algebraic decoders are presented, by the authors, for
the binary QR codes of lengths 71, 79, 89, 97, 103, and 113,
respectively
\cite{IEEEhowto:Chang2003}-\cite{IEEEhowto:Truong103}.

A strict requirement to apply the original Feng's or the modified
He's algorithms is that the weight of error pattern must be given.
This leads to step-by-step decoding algorithms and then the error
locator polynomial may not be determined in one step. Moreover,
when the code length increases, the desired matrices do not exist
in high-weight error cases. For example, no single matrix can be
used in the weight-5 error case of the QR code with length 47;
actually, He et al. used two matrices rather than one. To decode
each of the six QR codes of lengths 71 through 113, the authors
\cite{IEEEhowto:Chang2003}-\cite{IEEEhowto:Truong103} used at
least two matrices.

In order to develop one-step decoders, the Lagrange interpolation
formula (LIF) was applied \cite{IEEEhowto:Chang2010} to yield
representations of the unknown syndromes in terms of known
syndromes when the codes are generated by irreducible polynomials.
The LIF is a well known technique. For any finite set of points in
the $xy$-plane with all the $x$-coordinates distinct, the LIF can
be used to provide a polynomial function whose graph passing
through these given points. The LIF can be applied in finite
fields. And, in the unknown syndrome representation case, the
obtained polynomial has nice properties: All its coefficients are
coming from the base field; moreover, the degree of each term has
the same remainder when divided by the code length, as mentioned
in the following theorem appeared in \cite{IEEEhowto:Chang2010}.

{\it Theorem A.} (LIF) Let $\Omega =\{e(\beta) \mid 1\leq wt(e(x))
\leq t\}$ be the set of all correctable syndromes with $t$ the
error correcting capacity. For $a=\beta^{l_1}+\cdots+\beta^{l_v}
\in \Omega$ with $0\leq l_1< \cdots <l_v \leq n-1$, $1 \leq v \leq
t$, let $a^{[r]}=\beta^{rl_1}+\cdots+\beta^{rl_i}$ with $1\leq r
\leq n-1$. Then the polynomial $L_r(x)$ defined below
\begin{eqnarray}
L_r(x)=\sum_{a\in \Omega} \frac{a^{[r]}}{H'(a)} \frac{H(x)}{x-a},
\end{eqnarray}
where $H(x)=\prod_{a\in \Omega} (x-a)$ and $H'(x)$ the derivative
of $H(x)$, has the property $L_r(a)=a^{[r]}$ for all $a\in \Omega$
and is of the form $L_r(x)=x^r \varphi(x^n)$ for some
$\varphi(x)\in {\Bbb F_2[x]}$.

Actually, Theorem A is an application of the following theorem
which is also proved in \cite{IEEEhowto:Chang2010}.

{\it Theorem B:} Let $t$ be the error-correcting capacity and
$f\in \Bbb F_2[x_1,\ldots,x_t]$ be a homogeneous symmetric
function of degree $r$. Then there is a polynomial $A\in \Bbb
F_2[x]$ such that
\begin{eqnarray}
f(z_1,\ldots,z_t)=A((z_1+\cdots+z_t)^n)(z_1+\cdots+z_t)^r
\end{eqnarray}
for all $(z_1,\ldots,z_t)\in
T_t=\{(\beta^{l_1},\ldots,\beta^{l_v},0,\ldots,0)\mid
0\leq{l_1}<\cdots
\leq{l_v}\leq{n-1}\mbox{~and~}1\leq{v}\leq{t}\}$.

From Theorem A, the gap of the consecutive syndrome sequence can
be filled in and then the binary cyclic codes generated by
irreducible polynomials can be decoded by the BM algorithm.

On the other hand, when the generator polynomial is not
irreducible, the unknown syndromes cannot be expressed as a
univariate function of any known syndrome. What we need here is a
multivariate function. However, there is no, known to authors,
multivariate interpolation formula (MVIF) over finite fields. In
this paper, one such MVIF is developed based on the proposed
Kronecker delta function over finite field. It is also proved that
when applied to homogeneous symmetric rational functions on a
certain set, the obtained polynomial has significant properties;
namely, all the coefficients come from the base field and the
degree of each term satisfies a congruence equation with modulus
the code length. This is quite similar to that in the univariate
case mentioned in Theorem B. Next, based on the proposed MVIF, the
unknown syndrome representation method mentioned in
\cite{IEEEhowto:Chang2010} can be modified and then applied to any
cyclic codes, which will be described in Section \ref{AppUnkSyn}.

Besides the unknown syndrome representation method, Orsini and
Sala \cite{IEEEhowto:Sala2005} presented an algebraic decoder
based on the general error locator polynomial provided by the
Gr{\"o}bner basis method, and for any cyclic code the correctable
error patterns can be determined in one step. For those binary
cyclic codes generated by irreducible polynomials, the general
error locator polynomials can also be produced by applying the
Lagrange interpolation formula \cite{IEEEhowto:Chang2010}.

Furthermore, Orsini/Sala \cite{IEEEhowto:Orsini2007} and Augot et
al. \cite{IEEEhowto:Augot2009}, respectively, dealt with cyclic
codes generated by reducible polynomials, in which the
coefficients of error locator polynomials depending on more than
one known syndromes.

\subsection{Motivation}
\label{Motivation}

When decoding the binary quadratic residue code of length 31 whose
generator polynomial is a product of three irreducible factors, we
obtained an explicit formula of Kronecker delta function and a
multivariate interpolation formula over the binary field $\Bbb
F_2$. The unknown syndrome representation for $S_3$ yielded from
this MVIF is a binary polynomial and the degree of each term
satisfies a congruent equation. Motivated by this example, a
general result for arbitrary finite fields is presented and
proved, and a one-step decoding algorithm is developed for any
cyclic codes.

\subsection{Main Results}
\label{Main Result} In what follows, the original contributions of
the paper are outlined as compared to the existing works.
\begin{enumerate}
\item An explicit Kronecker delta function is proposed over finite
field. For $a\in \Bbb E=\Bbb F_{q^m}$, if $N=q^m-1$,
$\delta_a(x)=1-x^N$ for $a=0$, and
$\delta_a(x)=-\sum_{i=1}^N(x/a)^i$, otherwise. \item A
multivariate interpolation formula (MVIF) is developed over finite
field. Given $M$ pairwise distinct vectors
$(x_1^1,\ldots,x_1^v),\ldots,(x_M^1,\ldots,x_M^v)$ in $\Bbb
E^v=\Bbb E\times \cdots \times \Bbb E$ and $M$ arbitrary elements
$y_1,\ldots,y_M$ in $\Bbb E$, the developed MVIF yields a
multivariate polynomial $y=L(x^1,\ldots,x^v)$ over $\Bbb E$ so
that for every $i\in \{1,\ldots,M\}$ one has
$y_i=L(x_i^1,\ldots,x_i^v)$. \item We prove that, on a certain
set, any homogeneous symmetric rational function can be expressed
as a polynomial function of some homogeneous symmetric rational
functions. This polynomial can be provided from the proposed MVIF
and has significant properties on both the coefficients and the
degrees. \item For any cyclic code, the unknown syndromes can be
expressed as polynomial functions of the known syndromes. The
polynomials can also be provided by the presented MVIF and also
have the same properties as mentioned in 3). \item By applying the
unknown syndrome representation and the BM algorithm, a one-step
decoding scheme is developed for any cyclic codes.
\end{enumerate}

\subsection{Organization} \label{Organization}
In Section \ref{MLIF}, all the theoretical results are presented,
including the Kronecker delta function as well as the multivariate
interpolation formula over finite fields and the relations among
homogeneous symmetric rational functions. The decoder and the
example based on the unknown syndrome representation are presented
in Section \ref{AppUnkSyn}. The proofs of Theorems 2 and 4 are
provided in Appendix.

\section{Multivariate Interpolation Formula}
\label{MLIF} Throughout this paper, let $n$ be a positive integer,
$q$ a prime power. If $m$ is the multiplicative order of $q$
modulo $n$, let $\Bbb F=\Bbb F_q$ and $\Bbb E=\Bbb F_{q^m}$ be the
base and extension fields of order $q$ and $q^m$, respectively.

\subsection{Kronecker Delta Function}
\label{Delta-Function} Since the set $\Bbb E^*$ of nonzero
elements in $\Bbb E$ forms a multiplicative group, if $N=|\Bbb
E^*|=q^m-1$, then $\gamma^N=1$ for $\gamma\in \Bbb E^*$, and if
$\gamma \in \Bbb E^*\backslash\{1\}$ then
$\gamma+\gamma^2+\cdots+\gamma^N=\gamma(1+\cdots+\gamma^{N-1})=\gamma((1-\gamma^N)/(1-\gamma))=0$.
Combining these two facts, one has an explicit Kronecker delta
function on $\Bbb E$.

{\bf Definition 1.} For $a\in \Bbb E$ and $N=|\Bbb E|-1$, let
$\delta_a(x)$ be the following function defined on $\Bbb E$.
\begin{eqnarray}
\delta_a(x)=\left\{\begin{array}{ll}1-x^N&\mbox{if $a=0$}\\
-\sum_{k=1}^N \left( \frac{x}{a} \right)^k &\mbox{if $a\ne
0.$}\end{array}\right. \nonumber
\end{eqnarray}

It is easy to see that the value of $\delta_a(x)$ is 1 if $x=a$
and 0 otherwise. Therefore, $\delta_a(x)$ can be viewed as a
finite field version of the Kronecker delta function on $\Bbb E$.
By using the function $\delta_a(x)$, a multivariate interpolation
formula is developed over finite fields in the next subsection.

\subsection{Multivariate
Interpolation Formula over Finite Fields} \label{} Based on the
proposed Kronecker delta function in Definition 1, one can develop
the interpolation formula directly.

{\bf Theorem 1.} (multivariate
interpolation formula over finite field)\\
Let $\Bbb E$ be a finite field and $N=|\Bbb E|-1$. For $M,~v$
positive integers, if
$\{(x_1^1,\ldots,x_1^v),\ldots,(x_M^1,\ldots,x_M^v)\}$ is a set of
$M$ pairwise distinct $v$-tuples in $\Bbb E^v=\Bbb E \times \cdots
\times \Bbb E$ and $y_1,\ldots, y_M$ are $M$ arbitrary elements in
$\Bbb E$, then the multivariate polynomial given below
\begin{eqnarray}
y=L(x^1,\ldots,x^v)=\sum_{i=1}^M y_i \prod_{j=1}^v
\delta_{x_i^j}(x^j) \nonumber
\end{eqnarray}
has the property $y_i=L(x_i^1,\ldots,x_i^v)$ for $i=1,\ldots,M$,
where, for $a\in \Bbb E$
\begin{eqnarray}
\delta_a(x)=\left\{\begin{array}{ll}1-x^N&\mbox{if $a=0$}\\
-\sum_{k=1}^N \left( \frac{x}{a} \right)^k &\mbox{if $a\ne
0.$}\end{array}\right. \nonumber
\end{eqnarray}

The MVIF presented in Theorem 1 can be applied to yield polynomial
relations among homogeneous symmetric rational functions and to
represent the unknown syndromes in terms of known syndromes for
arbitrary cyclic codes as mentioned in Theorems 2 and 4 below,
respectively.

\subsection{Homogeneous Symmetric Rational Functions}
\label{homosymm}

In this subsection, we add one more condition on the parameters
$n$ and $q$; namely, gcd$(n, q-1) = 1$. Now, since $n$ divide
$q^m-1$, there exist primitive $n$th root of unity in
$\mathbb{E}$. Let $\beta$ be one of them. Then $\beta^i \neq 1$
for any $i \in \{1, \ldots,n-1\}$. With the additional condition
$gcd(n,q-1)=1$, one can prove that $\beta^i \notin \mathbb{F}^*$
for any $i \in \{1,\ldots,n-1\}$.

{\bf Lemma A.} If gcd$(n,q-1)=1$, then $\beta^i\notin \Bbb F^*$
for $i\in\{1,\ldots,n-1\}$.
\begin{IEEEproof}
If $\gamma=\beta^i\in \Bbb F^*$, then the order of $\gamma$ is
$k$, and by Lagrange theorem of finite group, $k$ divides $|\Bbb
F^*|=q-1$. Hence, $(\beta^i)^k=1,$ which implies $n\mid ik$. Since
$k\mid q-1$, gcd$(n,k)\mid \mbox{gcd}(n,q-1)$, and since
gcd$(n,q-1)=1$, gcd$(n,k)=1$. Finally, since $n\mid ik$ and
gcd$(n,k)=1$, one has $n\mid i$, which contradicts to the fact
$i<n$.
\end{IEEEproof}

Next, similar to the set $T_t$ defined in
\cite{IEEEhowto:Orsini2007}, \cite{IEEEhowto:Chang2010}, let
$T_t^*$, $t$ a positive integer, denote the following subset of
$\Bbb E^t$:
\begin{eqnarray}
T_t^*=\{(c_1\beta^{l_1},\ldots,c_v\beta^{l_v},0,\ldots,0) \mid
0\leq l_1 <\cdots < l_v< n, \nonumber\\
 c_1,\ldots,c_v\in \Bbb F^*,~1\leq v \leq t\}. \nonumber
\end{eqnarray}

Because gcd$(n,q-1)=1$, the elements of $T_t^*$ can be shown to
have unique representation.

{\bf Lemma B.} For $\vec{\theta},\vec{\theta'}\in T^*_t$ with
$\vec{\theta}=(c_1\beta^{l_1},\ldots,c_v\beta^{l_v},0,\ldots,0)$
and
$\vec{\theta'}=(c'_1\beta^{l'_1},\ldots,c'_1\beta^{l'_1},0,\ldots,0)$,
if $\vec{\theta}=\vec{\theta'}$ then $u=v$ and $c_i=c'_i$,
$l_i=l'_i$ for $i=1,\ldots,v$.
\begin{IEEEproof}
If $c_i\in \Bbb F^*$ and $l_i\in \{1,\ldots,n-1\}$ then
$c_i\beta^{l_i}\neq 0$ and then $u=v$. Next, if
$c_i\beta^{l_i}=c'_i\beta^{l'_i}$ with $c_i,c'_i\in \Bbb F^*$ and
$l_i,l'_i\in\{1,\ldots, n-1\}$ then
$\beta^{l_i-l'_i}=c_i^{-1}c'_i\in \Bbb F^*$. From Lemma A,
$l_i=l'_i$ and $c_i=c'_i$.
\end{IEEEproof}

Next, one defines some notations needed as follows. For
$\vec{\theta}=(c_1\beta^{u_1},\ldots,c_v\beta^{u_v},0,\ldots,0)\in
T^*_t$ and $w\in \{0,\ldots, n-1\}$, denote
\begin{eqnarray}
\label{proofeq1} \beta^w
\vec{\theta}=(c'_1\beta^{w_1},\ldots,c'_v\beta^{w_v},0,\ldots, 0),
\end{eqnarray}
where $w_1<\cdots<w_v$ with $w_i=v_{\pi(i)}+w~(\mbox{mod}~n)$ and
$c'_i=c_{\pi(i)}$ for some $\pi \in Sym(v)$ depending on $w$.
Similarly, for $w\in \{0,\ldots, m-1\}$, denote
\begin{eqnarray}
\label{proofeq2}
\vec{\theta}^{q^w}=(c'_1\beta^{w_1},\ldots,c'_v\beta^{w_v},0,\ldots,
0),
\end{eqnarray}
where $w_1<\cdots<w_v$ with $w_i=v_{\pi(i)} \times
q^w~(\mbox{mod}~n)$ and $c'_i=c_{\pi(i)}$ for some $\pi \in
Sym(v)$ depending on $w$. Note that for $\vec{\theta} \in T^*_t$,
the sum $c_1\beta^{l_1}+\cdots+c_v\beta^{l_v}$ is the first
syndrome $e(\beta)$ for the correctable error pattern
$e(x)=c_1x^{l_1}+\cdots+c_vx^{l_v}$. Moreover, if
$\beta^w\vec{\theta}= c'_1\beta^{w_1}+\cdots+c'_v\beta^{w_v}$ then
$c'_1\beta^{w_1}+\cdots+c'_v\beta^{w_v}$ is the syndrome of the
cyclic shift $x^we(x)$ and if $\vec{\theta}^{q^w} =
c'_1\beta^{w_1}+\cdots+c'_v\beta^{w_v}$ then
$c'_1\beta^{w_1}+\cdots+c'_v\beta^{w_v}$ is a conjugate of
$c_1\beta^{l_1}+\cdots+c_v\beta^{l_v}$ in $\Bbb E$. In addition,
since $\gamma^{q^m}=\gamma$ for $\gamma\in \Bbb E$,
$\vec{\eta}^{q^m}=\vec{\eta}$ for $\vec{\eta}\in T^*_t$. Hence,
there is a smallest positive integer $d$ denoted by
$deg(\vec{\theta})$ such that $\vec{\theta}^{q^d}=\vec{\theta}$.
Obviously, $deg(\vec{\theta})$ is a divisor of $m$.

Call a map $f$ from $T^*_t$ into $\Bbb E$ {\em $*$-homogeneous} of
degree $r$ if for $w \in \{0,\ldots,n-1\}$, $f
(\beta^w\vec{\theta})=(\beta^w)^r f(\vec{\theta})$. Also, $f$ is
called {\em $*$-symmetric} if for $w \in \{0, \ldots,m-1\}$,
$f(\vec{\theta}^{q^w})=(f(\vec{\theta}))^{q^w}$. It can be proved
that both the syndrome functions and the coefficient functions of
general error locator polynomial satisfy both conditions defined
above. Let $\mathcal{F}^*$ be the collection of all maps from
$T^*_t$ into $\Bbb E$ which are both $*$-homogeneous and
$*$-symmetric. For a positive integer $s$, $\{h_1, \ldots,h_s\}
\subset \mathcal{F}^*$ is said to be {\em $*$-independent} over
$T_t^*$ if the $s$-tuples $(h_1(\vec{\theta}),
\ldots,h_s(\vec{\theta}))$ are pairwise distinct for $\vec{\theta}
\in T^*_t$. A $*$-independent set plays the role as an algebraic
basis of $\mathcal{F}^*$, i.e., any map in $\mathcal{F}^*$ can be
expressed as a multivariate polynomial function in terms of the
elements from the $*$-independent set. Furthermore, the exponents
of each term in the multivariate polynomial satisfy a certain
condition, as described in the following theorem.

{\bf Theorem 2.} For positive integers $s,r_1,\ldots,r_s$, let
$h_1,\ldots,h_s\in \mathcal{F}^*$ be $*$-independent of
homogeneous degrees $r_1,\ldots,r_s$, respectively. Then for any
$f\in \mathcal{F}^*$ of homogeneous degree $r$, there is a
polynomial $L\in \Bbb F[x_1,\ldots,x_s]$ of the form
\begin{eqnarray}
L(x_1,\ldots,x_s)=\sum_{(i_1,\ldots,i_s)\in J}
c_{i_1,\ldots,i_s}x_1^{i_1}\cdots x_s^{i_s}, \nonumber
\end{eqnarray} where $J\subset \{(i_1,\ldots,i_s) \mid
r_1i_1+\cdots+r_si_s\equiv r \mbox{~(mod $n$)}\}$, such that
$f(\vec{\theta})=L(h_1(\vec{\theta}),\ldots,h_s(\vec{\theta}))$
for any $\vec{\theta}\in T^*_t$.

Note that, the obtained multivariate polynomial $L(x_1,...,x_s)$
is over the base field $\Bbb F$, that is, all the coefficients
$c_{i_1,\ldots,i_s}$'s are in $\Bbb F$, and moreover, the
exponents $i_1,\ldots,i_s$ of variables $x_i$'s in each term
$x_1^{i_1}\cdots x_s^{i_s}$ satisfy the congruence equation
$r_1i_1+\cdots+r_si_s \equiv r$ (mod $n$). It is easy to realize
that Theorem B (Theorem 1 in \cite{IEEEhowto:Chang2010}) is a
special case of Theorem 2 with $q=2$, $s=1$, $r_1=1$, and
$h(z_1,\ldots,z_t)=z_1+\cdots+z_t$.

\subsection{Application to Decode Cyclic Codes}
\label{AppCyCodes} Based on Theorems 1 and 2, algebraic decoders
for arbitrary cyclic codes can be developed. As above, let $n$ be
a positive integer, $q$ a prime power, with $gcd(n,q-1)=1$. And
let $m$ the multiplicative order of $q$ modulo $n$, $\Bbb F=\Bbb
F_q$, $\Bbb E=\Bbb F_{q^m}$, and $\beta \in \Bbb E$ a primitive
$n$th root of unity. Let $C$ be a cyclic code of length $n$ over
$\Bbb F$ generated by $g(x)=\prod_{i\in S_C} (x-\beta^i)$, where
$S_C$ is the {\it defining set} of $C$. Denote by $m_i(x)$ the
minimal polynomial of $\beta^i$ over $ \Bbb F_q$. If $R_C$ is a
subset of $S_C$ such that $g(x)=\prod_{i\in R_C} m_i(x)$ then call
$R_C$ a {\it base set} of $S_C$. Let $t$ be the error-correcting
capacity of $C$. An error pattern $e(x)$ is called {\it
correctable} if the number of nonzero terms in $e(x)\in \Bbb
E(x)$, i.e., $wt(e(x))$ or the weight of $e(x)$, is at most $t$.
Let $\mathcal E$ denote the set of all correctable error patterns.
For $i\in \{0,1,\ldots,n-1\}$, call $S_i=e(\beta^i)$ the $i$th
{\it syndrome} of the error pattern $e(x)$. If $i\in S_C$, $S_i$
can be calculated from the received word and is called the {\it
known syndrome} of $e(x)$; otherwise, for $i\notin S_C$, $S_i$ is
called the {\it unknown syndrome} of $e(x)$.

For $r\in\{0,\ldots,n-1\}$, define the syndrome map $S_r^*$ from
$T_t^*$ into $\Bbb E$ by
$S_r^*(c_1\beta^{l_1},\ldots,c_v\beta^{l_v},0,\ldots,0)=c_1\beta^{rl_1}+\cdots+c_v\beta^{rl_v}$.
Note that if $e(x)=c_1x^{l_1}+\cdots+c_vx^{l_v}\in \mathcal E$ and
$\vec{\theta}=(c_1\beta^{l_1},\ldots,c_v\beta^{l_v},0,\ldots,0)\in
T^*_t$, then
$S_r^*(\vec{\theta})=c_1\beta^{rl_1}+\cdots+c_v\beta^{rl_v}=e(\beta^r)=S_r$
is the $r$th syndrome of $e(x)$. Then, since elements of $T^*_t$
have unique representation, $S_r^*$ is well-defined and satisfies
both $*$-conditions mentioned above with homogeneous of degree
$r$.

{\bf Lemma C.} For $r \in \{0, \ldots, n-1 \}$, $S_r^* \in
\mathcal{F}^*$ with homogeneous degree $r$.

\begin{IEEEproof}
For $\vec{a}=\beta^w
\vec{\theta}=(c'_1\beta^{w_1},\ldots,c'_v\beta^{w_v},0,\dots,0)$,
where $w_i=l_{\pi(i)+w}$, $c'_i=c_{\pi(i)}$ for $i=1,\ldots,v$ and
$\pi\in Sym(v)$,
\begin{eqnarray}
S_r^*(\vec{a})&=&c'_1\beta^{rw_1}+\cdots+c'_v\beta^{rw_v}\nonumber\\
 &=&c_{\pi(1)}\beta^{r(l_{\pi(1)}+w)}+\cdots+c_{\pi(v)}\beta^{r(l_{\pi(v)}+w)}\nonumber\\
  &=&\beta^{rw}c_{\pi(1)}\beta^{rl_{\pi(1)}}+\cdots+\beta^{rw}c_{\pi(v)}\beta^{rl_{\pi(v)}}\nonumber\\
  &=&\beta^{rw}(c_{\pi(1)}\beta^{rl_{\pi(1)}}+\cdots+c_{\pi(v)}\beta^{rl_{\pi(v)}})\nonumber\\
  &=&\beta^{rw}S_r^*(\vec{\theta}).\nonumber
\end{eqnarray}
Hence, $S_r^*$ is $*$-homogeneous of degree $r$.

Next, for
$\vec{a}=\vec{\theta}^{q^w}=(c'_1\beta^{w_1},\ldots,c'_v\beta^{w_v},0,\ldots,0)$,
where $w_i=l_{\pi(i)}q^w$, $c'_i=c_{\pi(i)}$ for $i=1,\ldots,v$
and $\pi\in Sym(v)$,
\begin{eqnarray}
S_r^*(\vec{a})&=&c'_1\beta^{rw_1}+\cdots+c'_v\beta^{rw_v}\nonumber\\
 &=&c_{\pi(1)}\beta^{l_{\pi(1)}q^w}+\cdots+c_{\pi(v)}\beta^{l_{\pi(v)}q^w}\nonumber\\
  &=&\left(c_{\pi(1)}\beta^{l_{\pi(1)}}+\cdots+c_{\pi(v)}\beta^{l_{\pi(v)}}\right)^{q^w}\nonumber\\
  &=&\left(S_r^*(\vec{\theta})\right)^{q^w}.\nonumber
\end{eqnarray}
Hence, $S_r^*$ is $*$-symmetric and then belong to
$\mathcal{F}^*$.
\end{IEEEproof}

Let $s$ be the cardinality of $R_C$ and let
$R_C=\{r_1,\ldots,r_s\}$; consider the set of $s$-tuples
$\Omega=\{(e(\beta^{r_1}), \ldots, e(\beta^{r_s})) \mid e(x) \in
\mathcal E\}$. When $C$ is the binary quadratic residue code of
length 89, $R_C=\{1,5,9,11\}$, it has been proved
\cite{IEEEhowto:Shih89} that there is a one-to-one correspondence
between the sets $\mathcal E$ and
$\Omega=\{(e(\beta),e(\beta^5),e(\beta^9),e(\beta^{11}))\mid
e(x)\in \mathcal E\}$. This is actually true for any cyclic code.

{\bf Theorem 3.} For any cyclic code $C$, there is a one-to-one
correspondence between the set $\mathcal E$ of correctable error
patterns and the set $\Omega=\{(e(\beta^{r_1}), \ldots,
e(\beta^{r_s})) \mid e(x) \in \mathcal E\}$.
\begin{IEEEproof}
It suffices to prove that for $e_1(x),~e_2(x)\in \mathcal E$, one
has $e_1(x)=e_2(x)$ if and only if $(e_1(\beta^{r_1}), \ldots,
e_1(\beta^{r_s}))=(e_2(\beta^{r_1}), \ldots, e_2(\beta^{r_s}))$.
Since the ``necessary'' part is trivial, we only prove the
``sufficient'' part. If $(e_1(\beta^i):i\in
R_C)=(e_2(\beta^i):i\in R_C)$, then $e_1(\beta^i)-e_2(\beta^i)=0$
for any $i\in R_C$. Hence, for any $i\in R_C$, $m_i(x)$ divides
$e_1(x)-e_2(x)$ which implies that $g(x)$ divides $e_1(x)-e_2(x)$,
or equivalently, $e_1(x)-e_2(x)$ is a codeword, and then
$wt(e_1(x)-e_2(x))\geq d$. This leads to a contradiction because
$wt(e_1(x)-e_2(x))\leq wt(e_1(x))+wt(e_2(x))\leq 2t <d$.
\end{IEEEproof}

From Theorem 3, one has the following result immediately.

{\bf Theorem 4.} If $R_C=\{r_1,\ldots,r_s\}$, then
$\{S_{r_1}^*,\ldots,S_{r_s}^*\}$ is $*$-independent over $T_t^*$.

{\bf Corollary 1.} If $r\notin S_C$, then
$S_r^*=L(S_{r_1}^*,\ldots,S_{r_s}^*)$ and $S_r^*\in\mathcal{F}^*$.

From now on, for convenience, $T_t^*$ will be replaced by $T^*$
with $t$ the error correcting capacity. Note that, in the binary
case, i.e. $q=2$, there is a one-to-one correspondence between the
set $\mathcal E$ of correctable error patterns and the set $T$,
with $(\beta^{l_1},\ldots,\beta^{l_v},0,\ldots,0)\in T$
corresponding to the error patterns $e(x)=x^{l_1}+\cdots+x^{l_v}
\in \mathcal E$.

Next, for $i\in \{1,2,\ldots,n-1\}$ let
$x_i(z_1,\ldots,z_t)=z_1^i+\cdots+z_t^i \in \Bbb
F[z_1,\ldots,z_t]$ be the power symmetric polynomial of degree
$i$. If $\vec{\theta}=(\theta_1,\ldots,\theta_t)\in T$, and
$e(x)\in \mathcal E$ then
$x_i(\vec{\theta})=\theta_1^i+\cdots+\theta_t^i$ gives the $i$th
syndrome of $e(x)$, i.e. $x_i(\vec{\theta})=e(\beta^i)=S_i$. If
$R_C=\{r_1,\ldots,r_s\}$, then by Theorem 3 the power symmetric
polynomials $x_{r_1},\ldots,x_{r_s}$ satisfy the condition of
Theorem 2, i.e., $(x_{r_1}(\vec{\theta}),\ldots,
x_{r_s}(\vec{\theta}))$ are all distinct for $\vec{\theta}\in T$.
By Theorem 2 every homogeneous symmetric rational function can be
expressed as a polynomial in terms of $x_{r_1},\ldots,x_{r_s}$.
Since the unknown syndromes are also both homogeneous and
symmetric, they can be expressed as polynomials in terms of the
known syndromes $x_{r_1},\ldots,x_{r_s}$ and this unknown syndrome
representation is unified, i.e. independent of the occurred error
pattern $e(x)$. Hence, the decoding scheme in
\cite{IEEEhowto:Chang2010} based on the unified unknown syndrome
representation is valid for any binary cyclic codes. Furthermore,
for arbitrary prime power $q$, the unified unknown syndrome
representation method is also valid and will be shown in the
following section.

\section{Applications to Unknown Syndrome Representation}
\label{AppUnkSyn}
\subsection{Unified Unknown Syndrome Representation}
\label{} If $R_C=\{r_1,\ldots,r_s\}$ is the base set of $S_C$
mentioned above and $S_r$ is the unknown syndrome to be
determined, then by Theorem 1 there is a polynomial
$L(x_1,\ldots,x_s)$ over $\Bbb F$ such that
$S_r=L(S_{r_1},\ldots,S_{r_s})$. When $q=2$, since the syndromes
can be viewed as homogeneous symmetric functions on the set $T$,
by Theorem 2, the polynomial $L(x_1,\ldots,x_s)$ has the same form
as mentioned in Theorem 2. For the general case, $q$ is a prime
power with gcd$(n,q-1)=1$, this result is also true from Theorems
2 and 4 and as shown below.

{\bf Theorem 5.}\label{thm5} (unified unknown syndrome representation)\\
The unknown syndrome $S_r$ can be expressed as a polynomial
$L(S_{r_1},\ldots,S_{r_s})$ in terms of $S_{r_1},\ldots,S_{r_s}$
and the function $L\in \Bbb F[x_1,\ldots,x_s]$ is of the following
form
\begin{eqnarray}
L(x_1,\ldots,x_s)=\sum_{(i_1,\ldots,i_s)\in J}
c_{i_1,\ldots,i_s}x_1^{i_1}\cdots x_s^{i_s} \nonumber
\end{eqnarray}
where $J\subset \{(i_1,\ldots,i_s) \mid r_1i_1+\cdots+r_si_s\equiv
r_{s+1} \mbox{~(mod $n$)}\}$.

Since the proof of Theorem 5 is a slight modification of that of
Theorem 2, only the sketch proof is given in Appendix.

Now, with Theorem 5, the decoder based on the unknown syndrome
representation in \cite{IEEEhowto:Chang2010} can be modified and
applied to any cyclic codes as shown in next subsection.

\subsection{Decoding Algorithm} \label{DAsyn}

In this algorithm, we first calculate the known syndromes and use
them to determine the needed unknown syndromes to obtain $2t$
consecutive syndrome sequence. Next, apply the BM algorithm to
yield both the error locator polynomial $\sigma(x)$ and the error
evaluator polynomial $\Omega(x)$, and then for $v\leq t$, the
error pattern
$e(x)=e_{j_1}x^{j_1}+e_{j_2}x^{j_2}+\cdots+e_{j_{v}}x^{j_{v}}$ can
be determined by any appropriate root-position and -magnitude
calculations.

\noindent\hrulefill

\noindent{\bf Input: $r(x)$}
\begin{enumerate}
\item {Syndrome Calculation} 
\begin{enumerate} \item Known Syndrome Calculation\\
$~~~S_i=r(\beta^i),i\in R_C$.
\item Unknown Syndrome Calculation\\
for all $r\leq{2t}$, $r\notin S_C$,\\
$~S_r=L_r(S_{r_1},\ldots,S_{r_s}),$ for
$\{r_1,\ldots,r_s\}=R_C$,\\
$S_{r^q}=S_r^q.$
\end{enumerate}
\item {Error Locator and Evaluator Polynomials Calculation}\\
$\{\sigma(x),\Omega(x)\}=\mbox{IFBMA}(S_1,\ldots,S_{2t})$.
\item {Root-Position and -Magnitude Calculation}\\
(Apply the Chien search and Forney algorithm)\\
$e(x)=0$ \\
for $i$ from 0 to $n-1$ do\\
$~~~$if $\sigma(\beta^{-i})=0$ then $e_i=-\Omega({\beta^{-i}})/\sigma'^({\beta^{-i}})$\\
\end{enumerate}
{\bf Output:} $c(x)=r(x)-e(x)$

\noindent\hrulefill

IFBMA is the abbreviation for inverse-free Berlekamp-Massey
algorithm \cite{IEEEhowto:Chang2003}. To illustrate this decoding
algorithm, a workout example is given in next subsection.

\subsection{Decoding the (31, 16, 7) Binary Quadratic
Residue Code} \label{QR31Ex} Let $\Bbb E=\Bbb F_{2^5}$ be the
finite field of order 32 with multiplication modulus the primitive
polynomial $p(x)=1+x^2+x^5$. Since $(2^5-1)/31=1$, if $\alpha$ is
a root of $p(x)$ then $\alpha$ is also a primitive 31st root of
unity. Let $C$ be the triple-error-correcting (31, 16, 7) binary
QR code generated by the polynomial
$g(x)=1+x^3+x^8+x^9+x^{13}+x^{14}+x^{15}$. Then, $S_C =
\{1,2,4,5,7,8,9,10,14,16,18,19,20,25,28\}$ and the set
$R_C=\{1,5,7\}$ is a base set of $S_C$. Therefore, $S_1$, $S_5$,
$S_7$ are known syndromes, and $S_3$ is the needed unknown
syndrome for the application of the BM algorithm. From Theorem 2,
the unknown syndrome $S_3=L(S_1,S_5,S_7)$, where
$L(x,y,z)=\sum_{(i,j,k)\in J_1} x^iy^jz^k$ and $J_1$, listed in
Table I, contains 307 hexadecimal ordered triples. For instance,
the first triple (0,2,1E) in $J_1$, means the decimal ordered
triple (0,2,30) which indicates $y^2 z^{30}$ is a term of
$L(x,y,z)$. Since $C$ is a binary code, the only possible error
value is one and there is no error-magnitude calculation.

Now, suppose that the codeword $c(x)=0\in C$ is transmitted and a
corrupted codeword $r(x)=x^3+x^7+x^{20}$ is received, which
indicates a weight-3 error $e(x)=x^3+x^7+x^{20}$ occurred. Apply
the decoding algorithm of Section \ref{DAsyn} as shown below:
\begin{enumerate}
\item Calculate $2t$ consecutive syndromes:
\begin{enumerate}
\item Compute known syndromes:\\
$S_1=r(\alpha)=\alpha^3+\alpha^7+\alpha^{20}=\alpha^4,$\\
$S_2=S_1^2=\alpha^8,$\\ $S_4=S_2^2=\alpha^{16},$\\
$S_5=r(\alpha^5)=(\alpha^5)^{3}+(\alpha^5)^{7}+(\alpha^5)^{20}=\alpha^{16},$\\
$S_7=r(\alpha^7)=(\alpha^7)^{3}+(\alpha^7)^{7}+(\alpha^7)^{20}=0.$
\item Compute unknown syndromes:\\
$S_3=L(S_1,S_5,S_7)=\sum_{(i,j,k)\in J_1} S_1^i\times S_5^j\times
S_7^k$\\
$~~~~=S_1^0\times S_5^2\times
S_7^{30}+\cdots+S_1^{31}\times S_5^{13}\times S_7^0$\\
$~~~~=\alpha^{27},$\\
$S_6=S_3^2=\alpha^{54}=\alpha^{23}.$
\end{enumerate}
\item Apply IFBMA to obtain the
error locator polynomial:\\
The inputs are the six consecutive syndromes $S_1,S_2,\ldots,S_6$.
While $k=0$, set $E^{0}(x)=1$, $A^{0}(x)=1$, $l^{(0)}=0$, and
$\gamma^{(0)}=1$. Then apply IFBMA, as shown in Table
\ref{tableII}, to obtain the error locator polynomial
$E^{(6)}(x)=\alpha^{12}+\alpha^{16}x+\alpha^3x^2+\alpha^{11}x^3$
in the 6th procedure. \item Use Chien search or other method to
find the error
locations, and then correct the obtained errors.\\
The three roots, $\alpha^{11}$, $\alpha^{24}$, $\alpha^{28}$, of
$E^{(6)}(x)$ can be figured out and then the error locations are
$-11\equiv 20$, $-24\equiv 7$, and $-28\equiv 3~($mod$~31)$.
Finally, the recovered codeword is
$c(x)=r(x)-e(x)=(x^3+x^7+x^{20})-(x^3+x^7+x^{20})=0$ which is as
desired.
\end{enumerate}

\section{Applications to General Error Locator Polynomial}
\label{AppGELP} The notion of general error locator polynomials
was introduced by Orsini and Sala in
\cite{IEEEhowto:Sala2005,IEEEhowto:Orsini2007}, which is based on
the following two facts: First, each coefficient of the error
locator polynomial can be expressed as a function of the known
syndromes. Next, there are one-to-one correspondences among the
set of correctable error patterns, the set of known syndromes, and
the set of error locator polynomials. The general error locator
polynomial can then be defined in terms of the known syndromes and
can be used to obtain the error locator polynomial without
applying the Berlekamp-Massey method or any others. Orsini and
Sala use the Gr{\"o}bner basis method to determine the general
error locator polynomial for arbitrary cyclic codes by calculating
each of its coefficients separately.

In this paper, we apply the MVIF instead of the Gr{\"o}bner basis
method to determine the general error locator polynomial also for
arbitrary cyclic codes. Moreover, we prove that the degrees of
each terms in the obtained polynomials satisfy a congruence
condition.

\subsection{General Error Locator Polynomial} \label{GELP}
For $i\in\{0,\ldots,t\}$, define the map $\sigma_i^*$ from $T_t^*$
into $\Bbb E$ by
$\sigma_i^*(c_1\beta^{l_1},\ldots,c_v\beta^{l_v},0,\ldots,0)=\sum_{k_1<\cdots<k_i}
\prod_{j=1}^i \beta^{l_{k_j}}$. Then, since elements of $T^*_t$
have unique representation, $\sigma_i^*$ is well-defined and
satisfies both $*$-conditions mentioned in Section \ref{homosymm}
with homogeneous of degree $i$.

{\bf Lemma D.} For $i \in \{0, \ldots, t \}$, $\sigma_r^* \in
\mathcal{F}^*$ with homogeneous degree $i$.

\begin{IEEEproof}
For $\vec{a}=\beta^w
\vec{\theta}=(c'_1\beta^{w_1},\ldots,c'_v\beta^{w_v},0,\dots,0)$,
where $w_i=l_{\pi(i)+w}$, $c'_i=c_{\pi(i)}$ for $i=1,\ldots,v$ and
$\pi\in Sym(v)$,
\begin{eqnarray}
\sigma_i^*(\vec{a})&=&\sum_{k_1<\cdots<k_i} \prod_{j=1}^i
\beta^{w_{k_j}}\nonumber\\
 &=&\sum_{k_1<\cdots<k_i} \prod_{j=1}^i
\beta^{l_{\pi(k_j)}+w}\nonumber\\
 &=&\sum_{k_1<\cdots<k_i} \beta^{iw} \prod_{j=1}^i
\beta^{l_{\pi(k_j)}}\nonumber\\
 &=&\beta^{iw} \sum_{k_1<\cdots<k_i} \prod_{j=1}^i
\beta^{l_{\pi(k_j)}}\nonumber\\
 &=&\beta^{iw}\sigma_i^*(\vec{\theta}).\nonumber
\end{eqnarray}
Hence, $\sigma_i^*$ is $*$-homogeneous of degree $i$.

Next, recall that for
$\vec{a}=\vec{\theta}^{q^w}=(c'_1\beta^{w_1},\ldots,c'_v\beta^{w_v},0,\ldots,0)$,
where $w_i=l_{\pi(i)}q^w$, $c'_i=c_{\pi(i)}$ for $i=1,\ldots,v$
and $\pi\in Sym(v)$,
\begin{eqnarray}
\sigma_i^*(\vec{a})&=&\sum_{k_1<\cdots<k_i} \prod_{j=1}^i
\beta^{w_{k_j}}\nonumber\\
 &=&\sum_{k_1<\cdots<k_i} \prod_{j=1}^i
\beta^{l_{\pi(k_j)}q^w}\nonumber\\
 &=&\sum_{k_1<\cdots<k_i} \left(\prod_{j=1}^i
\beta^{l_{\pi(k_j)}}\right)^{q^w}\nonumber\\
 &=&\left(\sum_{k_1<\cdots<k_i} \prod_{j=1}^i
\beta^{l_{\pi(k_j)}}\right)^{q^w}\nonumber\\
 &=&\left(\sigma_i^*(\vec{\theta})\right)^{q^w}.\nonumber
\end{eqnarray}
Hence, $\sigma_i^*$ is $*$-symmetric and then belong to
$\mathcal{F}^*$.
\end{IEEEproof}

For
$\vec{\theta}=(c_1\beta^{l_1},\ldots,c_v\beta^{l_v},0,\ldots,0)\in
T_t^*$, let $\tilde \sigma(\vec{\theta},z)=(z-\beta^{l_1})\cdots
(z-\beta^{l_v})$. Then $\tilde \sigma(\vec{\theta},z)=z^t+\tilde
\sigma_{t-1}(\vec{\theta})z^{t-1}+\cdots+\tilde
\sigma_1(\vec{\theta})z+\tilde \sigma_0(\vec{\theta})$, where
$\tilde \sigma_i(\vec{\theta})=(-1)^{t-i}\sum_{k_1<\cdots<k_i}
\prod_{j=1}^i \beta^{l_{k_j}}$ for $i=0,\ldots,t-1$. Note that,
from Lemma D, each $\tilde \sigma_i$ belongs to $\mathcal F^*$ and
is of homogeneous degree $i$, and by Theorems 2 and 4 can be
expressed as a multivariate polynomial in terms of the known
syndrome functions $S_{r_1}^*,\ldots,S_{r_s}^*$, where
$R_C=\{r_1,\ldots, r_s\}$. That is, $\tilde
\sigma_i(\vec{\theta})=\tilde
\sigma_i^*(x_1(\vec{\theta}),\ldots,x_s(\vec{\theta}))$, where
$x_1=S_{r_1}^*,\ldots,x_s=S_{r_s}^*$ and
$\sigma_i^*(x_1,\ldots,x_s)\in \Bbb F[x_1,\ldots,x_s]$. The
polynomial
$\sigma^*(X,z)=z^t+\sigma_{t-1}^*(X)z^{t-1}+\cdots+\sigma_1^*(X)z+\sigma_0^*(X)$
with $X=(x_1,\ldots,x_s)$ is called the {\it general error locator
polynomial} of the cyclic code $C$.

For any cyclic code, the coefficients $\sigma_i^*(x_1,\ldots,x_s)$
defined above has the following form, as a consequence of Theorem
2.

{\bf Theorem 6.} \label{thm6} (Coefficients of general error locator polynomial)\\
For $0\leq{i}\leq{t-1}$, the coefficient
$\sigma_i^*(x_1,\ldots,x_s)$ of the general error locator
polynomial $\sigma^*(x_1,\ldots,x_s,z)$ is in $\Bbb
F[x_1,\ldots,x_s,z]$ and is of the form
\begin{eqnarray}
\sigma_i^*(x_1,\ldots,x_s)=\sum_{(i_1,\ldots,i_s)\in J}
c_{i_1,\ldots,i_s} x_1^{i_1}\cdots x_s^{i_s} \nonumber
\end{eqnarray}
where $J$ is a subset of $\{(i_1,\ldots,i_s) \mid
r_1i_1+\cdots+r_si_s\equiv r_{s+1} \mbox{~(mod $n$)}\}$.

\subsection{Decoding Algorithm}
\label{DAgelp} To decode a corrupted data, we first calculate the
known syndromes and use them to determine the coefficients of the
general error locator polynomial. The obtained polynomial is
actually the error locator polynomial. With some appropriate root
position and magnitude calculations, the error can be pointed out
and the original data can be recovered.

\noindent\hrulefill

\noindent{\bf Input: $r(x)$}
\begin{enumerate}
\item {Syndrome Calculation}\\ 
for all $i\leq 2t,~S_i=r(\beta^i)$\\
\item {General Error Locator Polynomial Calculation}\\
$\sigma(x)=\sigma(X,x),~X=(S_{k_1},\cdots,S_{k_v}),~k_1,\ldots,k_v\in
S_C\cap R$. \item {Root-Finding and Magnitude Calculation}\\
$\Omega(x)\equiv S(x)\times \sigma(x)$ (mod $x^{2t}$)\\
$e(x)=0$\\
for $i$ from 0 to $n-1$ do\\
$~~~$if $\sigma(\beta^{-i})=0$ then $e_i=-\Omega({\beta^{-i}})/\sigma'({\beta^{-i}})$\\

\end{enumerate}
{\bf Output:} $c(x)=r(x)-e(x)$

\noindent\hrulefill

\subsection{Decoding the (15,11,5) RS Code}
\label{RSex} The decoding scheme based on the general error
locator polynomial is illustrated in the following example. Let
$\Bbb E=\Bbb F_{2^4}$ be the finite field of order 16 with the
multiplication modulus the primitive polynomial $p(x)=1+x+x^4$. If
$\alpha$ is the root of $p(x)$ then $\alpha$ is also a primitive
15th root of unity. Let $C$ be the double-error-correcting (15,
11, 5) RS code generated by the polynomial $g(x)=\prod_{i=1}^4
(x+\alpha^i)=\alpha^{13}+\alpha^6 x+\alpha^3 x^2+\alpha^{10}
x^3+x^4$. For the known syndromes $S_1$, $S_2$, $S_3$, and $S_4$,
let the general error locator polynomial be
$\sigma(X,x)=1+\tilde\sigma_1(X)x+\tilde\sigma_2(X)x^2$, which
$X=(S_1,S_2,S_3,S_4)$ and $\tilde \sigma_1(X)=\sum_{(i,j,k,u)\in
A_2} x_1^ix_2^jx_3^kx_4^u$, $\tilde \sigma_2(X)=\sum_{(i,j,k,u)\in
A_3} x_1^ix_2^jx_3^kx_4^u$, where $A_2$ and $A_3$ are shown in
Table \ref{tableIII} and Table \ref{tableIV}.

Suppose a zero codeword $c(x)={\mathbf 0}\in C$ is transmitted,
and the received corrupted codeword
$r(x)=\alpha^6x^2+\alpha^5x^{14}$, which indicates that the error
locations are 2 and 14, and the corresponding error values are
$\alpha^5$ and $\alpha^6$. Apply the decoding algorithm of Section
\ref{DAgelp} as the following steps.
\begin{enumerate}
\item Syndromes calculation:\\
$S_i=r(\alpha^i)=\alpha^6\times ({\alpha^i})^2+\alpha^5\times
({\alpha^i})^{14}$ where $i=1,2,3,4.$
$S_1=\alpha^5,S_2=\alpha^{12},S_3=\alpha^7$, and $S_4=\alpha^7$.\\
$S(x)=\alpha^5+\alpha^{12}x+\alpha^7x^2+\alpha^7x^3$

\item General error locator polynomial calculation:\\
$\tilde \sigma_1(X)=\sum_{(i,j,k,u)\in A_2} S_1^i\times S_2^j
\times
S_3^k\times S_4^u$\\
$~~~~~~~~~~=S_1^0\times S_2^2\times S_3^{10}\times S_4^3+\cdots+S_1^{15}\times S_2^{14}\times S_3^1\times S_4^0$\\
$~~~~~~~~=\alpha^{13}$\\ and \\ $\tilde
\sigma_2(X)=\sum_{(i,j,k,u)\in A_3} S_1^i\times S_2^j \times
S_3^k\times S_4^u$\\
$~~~~~~~~~~~=S_1^2\times S_2^9\times S_3^4\times S_4^{15}+\cdots+S_1^{15}\times S_2^{15}\times S_3^{13}\times S_4^2$\\
$~~~~~~~~=\alpha.$\\
Therefore,
$\sigma(X,x)=1+\sigma_1(X)x+\sigma_2(X)x^2=1+\alpha^{13}x+\alpha
x^2$.

\item Root-finding and magnitude calculation:\\
The two polynomials $\sigma(x)$ and $\Omega(x)$ satisfy the degree
relation, deg $\Omega(x)<$ deg $\sigma(x)=2$, and $\Omega(x)\equiv
S(x)\times \sigma(x)$ (mod $x^{4}$). Therefore, set
$\Omega(x)=\alpha^6+\alpha^8x$ and $\sigma'(x)=\alpha^{13}$. One
can find two roots of $\sigma(x)$ are $\alpha^1$ and
$\alpha^{13}$, which indicates that the locations of errors are 2
and 14. So $e_2=-\Omega(\alpha^{13})/\alpha^{13}=\alpha^6$ and
$e_{14}=-\Omega(\alpha^{1})/\alpha^{13}=\alpha^5$. The error
polynomial is $e(x)=\alpha^6 x^2+\alpha^5 x^{14}$. Finally, the
recovered codeword equals $c(x)=r(x)-e(x)=0$, and finish the
decoding algorithm.
\end{enumerate}

\section{Concluding Remark}
\label{Conclusion} The polynomials obtained from the proposed MVIF
may contain quite large number of terms. However, in the decoding
application, the situation may be mitigated. Actually, efforts
have been made to reduce the number of terms in the unknown
syndrome representation. One example is, by adding some extra
points during the applications of MVIF to determine the unified
unknown syndrome representation of the binary quadratic residue
code of length 41, one may deduce the number of terms from 1355 in
the original obtained polynomial to 1295 in a new obtained
polynomial, and these two polynomials are equivalent, i.e. have
the same values on the set $\Omega$ of known syndromes. This is
just a first stage trial; that is, we just add the fewest possible
points needed to yield a new polynomial in each MVIF calculation.
We believe that some better polynomials, from the viewpoint of
hardware design, will be obtained by adding more extra points in
each MVIF calculation.

\appendix
\subsection{Proof of Theorem 2}

Since all the vectors
$(h_1(\vec{\theta}),\ldots,h_s(\vec{\theta}))$, $\vec{\theta}\in
T^*_t$, are pairwise distinct, by Theorem 1, there is a polynomial
$L\in \Bbb E(x_1,\ldots,x_s)$ such that
$f(\vec{\theta})=L(h_1(\vec{\theta}),\ldots,h_s(\vec{\theta}))$
for $\vec{\theta}\in T^*_t$, and $L(x_1,\ldots,x_s)$ is of the
form
\begin{eqnarray}
L(x_1,\ldots,x_s)=\sum_{\vec{\theta}\in T^*_t} f(\vec{\theta})
\prod_{i=1}^s \delta_{h_i(\vec{\theta})}(x_i), \nonumber
\end{eqnarray}
where $\delta_a(x)$ is the Kronecker delta function defined in
Definition 1. The polynomial $L(x_1,\ldots,x_s)$  above can be
written as
\begin{eqnarray}
L(x_1,\ldots,x_s)=\sum_{(i_1,\ldots,i_s)\in J} c_{i_1,\ldots,i_s}
x_1^{i_1}\cdots x_s^{i_s}, \nonumber
\end{eqnarray}
where $c_{i_1,\ldots,i_s} \in \Bbb E$ and $J\subset
\{(i_1,\ldots,i_s) \mid i_1,\ldots,i_s\in \{0,1,2,\ldots\}\}$. To
prove the theorem, it suffices to show the following two claims:
\begin{enumerate}
\item $c_{i_1,\ldots,i_s} \in \Bbb F$ for $(i_1,\ldots,i_s)\in J$,
and \item $J\subset \{(i_1,\ldots,i_s) \mid
r_1i_1+\cdots+r_si_s\equiv r (\mbox{mod }n)\}$.
\end{enumerate}

Next, let ``$\sim$'' and ``$\approx$'' be two relations on $T^*_t$
defined by
\begin{eqnarray}
\label{proofeq3} \vec{\theta}\sim \vec{\theta}' \mbox{ if }
\vec{\theta}'=\beta^w\vec{\theta}
\end{eqnarray}
for some $w\in \{0, \ldots, n-1\}$ and
\begin{eqnarray}
\label{proofeq4} \vec{\theta}\approx \vec{\theta}'' \mbox{ if }
\vec{\theta}''={\vec{\theta}}^{q^w}
\end{eqnarray}
for some $w\in \{0, \ldots, m-1\}$. Then, obviously, both
``$\sim$'' and ``$\approx$'' are equivalence relations on $T^*_t$.

For $\vec{\theta}\in T^*_t$ and $d=deg(\vec{\theta})$, denote by
$\langle \vec{\theta} \rangle$ and $[\vec{\theta}]$ the
equivalence classes of $\vec{\theta}$ with respect to ``$\sim$''
and ``$\approx$'', respectively. Then one has
\begin{eqnarray}
\label{proofeq5} \langle \vec{\theta} \rangle=\{ \vec{\theta}'
\mid \vec{\theta} \sim \vec{\theta}'\}=\{\beta^w \vec{\theta} \mid
w=0,\ldots,n-1\}
\end{eqnarray}
and
\begin{eqnarray}
\label{proofeq6} [\vec{\theta}]=\{ \vec{\theta}'' \mid
\vec{\theta} \approx \vec{\theta}''\}=\{ \vec{\theta}^{q^w} \mid
w=0,\ldots,d-1\}.
\end{eqnarray}
Let $R_1$ (resp. $R_2$) be a complete set of representatives of
``$\sim$'' (resp.``$\approx$'') in $T^*_t$, i.e. $T^*_t$ can be
expressed as a disjoint union of $\langle \vec{\theta} \rangle$
(resp. $[\vec{\theta}]$) for $\vec{\theta}\in R_1$ (resp.
$\vec{\theta}\in R_2$), or equivalently,
$T^*_t=\bigcup_{\vec{\theta}\in R_1} \langle \vec{\theta} \rangle$
(resp. $T^*_t=\bigcup_{\vec{\theta}\in R_2} [\vec{\theta}]$).
Hence, the polynomial $L(x_1,\ldots,x_s)$ can be expressed as
either
\begin{eqnarray}
\sum_{\vec{\theta}\in R_1} \sum_{\vec{a}\in \langle \vec{\theta}
\rangle} f(\vec{a}) \prod_{i=1}^s \delta_{h_i(\vec{a})} (x_i)
\nonumber
\end{eqnarray}
or
\begin{eqnarray}
\sum_{\vec{\theta}\in R_2} \sum_{\vec{a}\in [\vec{\theta}]}
f(\vec{a}) \prod_{i=1}^s \delta_{h_i(\vec{a})} (x_i). \nonumber
\end{eqnarray}
We will prove that for every $\vec{\theta} \in R_1$ Claim (1) is
true for the polynomial
\begin{eqnarray}
L_{\langle \vec{\theta} \rangle}(x_1,\ldots,x_s)= \sum_{\vec{a}\in
\langle \vec{\theta} \rangle} f(\vec{a}) \prod_{i=1}^s
\delta_{h_i(\vec{a})} (x_i) \nonumber
\end{eqnarray}
and for $\vec{\theta} \in R_2$ Claim (2) is true for the
polynomial
\begin{eqnarray}
L_{[\vec{\theta}]}(x_1,\ldots,x_s)= \sum_{\vec{a}\in
[\vec{\theta}]} f(\vec{a}) \prod_{i=1}^s \delta_{h_i(\vec{a})}
(x_i), \nonumber
\end{eqnarray}
which will imply that both Claims (1) and (2) hold for the
polynomial $L(x_1,\ldots,x_s)$.

For
$\vec{\theta}=(c_1\beta^{l_1},\ldots,c_v\beta^{l_v},0,\ldots,0)\in
T^*_t$ and $\vec{a}=\beta^w \vec{\theta}
=(c'_1\beta^{w_1},\ldots,c'_v\beta^{w_v},0,\ldots,0)\in \langle
\vec{\theta} \rangle$, if $h \in \mathcal{F}^*$ has
$*$-homogeneous degree $r$, then
$h(\vec{a})=h(\beta^w\vec{\theta})=h(c_1\beta^{l_1},\ldots,c_v\beta^{l_v},0,\ldots,0)=\beta^{wr}h(c_1\beta^{u_1},\ldots,c_v\beta^{u_v},0,\ldots,0)=(\beta^w)^rh(\vec{\theta})=\beta^{rw}h(\vec{\theta})$,
since for each $i$, $w_i \equiv w+u_j$ (mod $n$) for some $j$.
Similarly, for $\vec{a}=\vec{\theta}^{q^w}\in [\vec{\theta}]$, if
$h(z_1,\ldots,z_t)$ is a $*$-symmetric, then
$h(\vec{a})=h(\vec{\theta}^{q^w})=(h(\vec{\theta}))^{q^w}$. Hence,
for either $\vec{a} \in \langle \vec{\theta} \rangle$ or
$\vec{a}\in [\vec{\theta}]$, $h_i(\vec{\theta})=0$ if and only if
$h_i(\vec{a})=0$ for $i=1,\ldots,s$.

Next, let $U_0$ and $U_1$ be subsets of $\{1,\ldots, s\}$ defined
below:
\begin{eqnarray}
U_0=U_0(\vec{\theta})=\{i \mid h_i(\vec{\theta})=0\}\nonumber
\end{eqnarray}
and
\begin{eqnarray} U_1=U_1(\vec{\theta})=\{1,\ldots,s\}\setminus U_0.
\nonumber
\end{eqnarray}
Let $A$ be either $\langle \vec{\theta} \rangle$ or
$[\vec{\theta}]$. The polynomial $L_A(x_1,\ldots,x_s)$ can be
written as, from the definition of the Kronecker delta-function,
\begin{eqnarray}
L_A(x_1,\ldots,x_s)= \sum_{\vec{a}\in A} f(\vec{a}) \prod_{i\in
U_0} \delta_{h_i(\vec{a})} (x_i) \prod_{i\in U_1}
\delta_{h_i(\vec{a})} (x_i) \nonumber\\
= \sum_{\vec{a}\in A} f(\vec{a}) \prod_{i\in U_0} (1-x_i^N)
\prod_{i\in U_1} (-1)
\sum_{k=1}^N\left(\frac{x_i}{h_i(\vec{a})}\right)^k
\nonumber\\
=\prod_{i\in U_0} (1-x_i^N) \sum_{\vec{a}\in A} f(\vec{a})
\prod_{i\in U_1} (-1)
\sum_{k=1}^N\left(\frac{x_i}{h_i(\vec{a})}\right)^k
\nonumber\\
=\prod_{i\in U_0} (1-x_i^N) L_A^*(x_1,\ldots,x_s),
\nonumber\\
\end{eqnarray}
where $L_A^*(x_1,\ldots,x_s)=\sum_{\vec{a}\in A} f(\vec{a})
\prod_{i\in U_1} (-1) \sum_{k=1}^N(\frac{x_i}{h_i(\vec{a})})^k$.
Since $\prod_{i\in U_0} (1-x_i^N)$ is a polynomial over the base
field $\Bbb F$, if $L_A^*(x_1,\ldots,x_s)$ is a polynomial over
$\Bbb F$, then so does $L_A(x_1,\ldots,x_s)$. Moreover, since $n$
divides $N=q^m-1$, to show Claim (2) is true for
$L_A(x_1,\ldots,x_s)$, it suffices to show the same assertion for
$L_A^*(x_1,\ldots,x_s)$. Therefore, without loss of generality, we
may assume $L_A(x_1,\ldots,x_s)=L_A^*(x_1,\ldots,x_s)$ and
$U_1=\{1,\ldots,s\}$. That is,
\begin{eqnarray}
\label{proofeq7}
L_A(x_1,\ldots,x_s)= \sum_{\vec{a}\in A}
f(\vec{a}) \prod_{i=1}^s (-1) \sum_{k=1}^N
\left(\frac{x_i}{h_i(\vec{a})}\right)^k
\nonumber\\
=(-1)^s \sum_{\vec{a}\in A} f(\vec{a})  \sum_{k_1,\ldots,k_s}
\prod_{i=1}^s \left(\frac{x_i}{h_i(\vec{a})}\right)^{k_i}
\nonumber\\
=(-1)^s \sum_{k_1,\ldots,k_s} \left(\sum_{\vec{a}\in A}
\frac{f(\vec{a})}{(h_1(\vec{a}))^{k_1}\cdots
(h_s(\vec{a}))^{k_s}}\right) x_1^{k_1}\cdots x_s^{k_s}.
\end{eqnarray}

To prove Claim (1), let $A=[\vec{\theta}]$ and show that
$\sum_{\vec{a}\in [\vec{\theta}]} f(\vec{a}) \prod_{i=1}^s
(h_i(\vec{a}))^{-k_i} \in \Bbb F$ for any index $(k_1,\ldots,
k_s)$. Since $deg(\vec{\theta})=d$,
$[\vec{\theta}]=\{\vec{\theta}^{q^w} \mid w=0, \ldots, d-1\}$, and
since $f,h_1,\ldots,h_s$ are $*$-symmetric, one has
\begin{eqnarray}
\sum_{\vec{a}\in [\vec{\theta}]}
\frac{f(\vec{a})}{(h_1(\vec{a}))^{k_1}\cdots
(h_s(\vec{a}))^{k_s}}=\sum_{w=0}^{d-1}
\frac{f(\vec{\theta}^{q^w})}{(h_1(\vec{\theta}^{q^w}))^{k_1}\cdots
(h_s(\vec{\theta}^{q^w}))^{k_s}} \nonumber\\
=\sum_{w=0}^{d-1}
\frac{(f(\vec{\theta}))^{q^w}}{((h_1(\vec{\theta}))^{k_1})^{q^w}\cdots
((h_s(\vec{\theta}))^{k_s})^{q^w}} \nonumber\\
=\sum_{w=0}^{d-1}
\left(\frac{f(\vec{\theta})}{(h_1(\vec{\theta}))^{k_1}\cdots
(h_s(\vec{\theta}))^{k_s}}\right)^{q^w} \nonumber\\
=\sum_{w=0}^{d-1} \gamma^{q^w}, \nonumber
\end{eqnarray}
where $\gamma={f(\vec{\theta})}{((h_1(\vec{\theta}))^{k_1}\cdots
(h_s(\vec{\theta}))^{k_s})^{-1}}\in \Bbb E$. Since
$\gamma^{q^d}={f(\vec{\theta}^{q^d})}{((h_1(\vec{\theta}^{q^d}))^{k_1}\cdots
(h_s(\vec{\theta}^{q^d}))^{k_s})^{-1}}={f(\vec{\theta})}{((h_1(\vec{\theta}))^{k_1}\cdots
(h_s(\vec{\theta}))^{k_s})^{-1}}=\gamma$, this implies $\gamma$
belongs to the subfield of $\Bbb E$ of dimension $d$ over $\Bbb
F$, and then $\sum_{w=0}^{d-1} \gamma^{q^w}\in \Bbb F.$

Next, let $A=\langle \vec{\theta} \rangle$. To prove Claim (2) is
equivalent to show that all the coefficients $\sum_{\vec{a}\in
\langle \vec{\theta} \rangle}
{f(\vec{\theta})}{((h_1(\vec{\theta}))^{k_1}\cdots
(h_s(\vec{\theta}))^{k_s})^{-1}}$ in (\ref{proofeq7}) are zero
except for those whose indices $(k_1,\ldots, k_s)$'s satisfying
the congruent equation $r_1k_1+\cdots+r_sk_s \equiv r$ (mod $n$).
Now, since the rational functions $f,h_1,\ldots,h_s$ are all
$*$-homogeneous of degrees $r,r_1,\ldots,r_s$, respectively, one
has
\begin{eqnarray}
\label{eq11} \sum_{\vec{a}\in \langle \vec{\theta} \rangle}
\frac{f(\vec{a})}{(h_1(\vec{a}))^{k_1}\cdots
(h_s(\vec{a}))^{k_s}}=\sum_{w=0}^{n-1} \frac{f(\beta^w
\vec{\theta})}{(h_1(\beta^w \vec{\theta}))^{k_1}\cdots
(h_s(\beta^w \vec{\theta}))^{k_s}} \nonumber\\
=\sum_{w=0}^{n-1}
\frac{\beta^{rw}f(\vec{\theta})}{\beta^{r_1wk_1}(h_1(\vec{\theta}))^{k_1}\cdots
\beta^{r_swk_s}(h_s(\vec{\theta}))^{k_s}} \nonumber\\
=\sum_{w=0}^{n-1} \beta^{w(r-r_1k_1-\cdots-r_sk_s)}
\frac{f(\vec{\theta})}{(h_1(\vec{\theta}))^{k_1}\cdots
(h_s(\vec{\theta}))^{k_s}} \nonumber\\
=\frac{f(\vec{\theta})}{(h_1(\vec{\theta}))^{k_1}\cdots
(h_s(\vec{\theta}))^{k_s}}\sum_{w=0}^{n-1}
\beta^{w(r-r_1k_1-\cdots-r_sk_s)}.
\end{eqnarray}
Since $\beta$ is a primitive $n$th root of unity,
$\sum_{w=0}^{n-1}\beta^{w\lambda}=0$ unless $\lambda$ is multiple
of $n$. The formula in (\ref{eq11}) equals zero, unless
$r-(r_1k_1+\cdots+r_sk_s)$ is a multiple of $n$, or equivalently,
$r_1k_1+\cdots+r_sk_s \equiv r$ (mod $n$). This completes the
proof of Claim 2 and then the proof of Theorem 2. \hfill
$\IEEEQEDclosed$

\begin{table*}
\renewcommand{\arraystretch}{1.2}
\begin{center}
\caption{List 307 triples $(i,j,k)$ of $J_1$ in hexadecimal
representation} \label{tableI}
\begin{tabular}{|c|c|c|c|c|c|c|c|c|c|c|c|c|} \hline
0,2,1E&0,3,10&0,4,2&0,5,13&0,6,5&0,8,8&0,9,19&0,A,B&0,E,11&0,10,14&0,12,17&
0,13,9&0,15,C\\\hline 0,1A,4&0,1B,15&0,1E,A&1,1,4&
1,8,1E&1,9,10&1,A,2&1,C,5&1,D,16&
1,E,8&1,10,B&1,14,11&1,17,6\\\hline 1,18,17&1,19,9&
1,1F,12&2,0,9&2,1,1A&2,2,C&2,3,1D&2,4,F&2,5,1&2,7,4&2,A,18&
2,D,D&2,F,10\\\hline 2,10,2&2,11,13&2,16,B&
2,18,E&2,19,0&2,1B,3&2,1F,9& 3,0,1F&3,1,11&3,3,14&3,6,9&3,7,1A&
3,8,C\\\hline 3,D,4&3,10,18&3,16,2&
3,1A,8&3,1C,B&4,2,19&4,5,E&4,7,11&
4,A,6&4,C,9&4,10,F&4,16,18&4,18,1B\\\hline

4,19,D&5,0,D&5,1,1E&5,2,10& 5,3,2&5,5,5&5,8,19&5,10,6&5,13,1A&
5,14,C&5,17,1&5,19,4&5,1C,18\\\hline

5,1D,A&6,0,4&6,6,D&6,8,10&6,9,2&
6,A,13&6,B,5&6,C,16&6,E,19&6,11,E&6,13,11&6,16,6&6,19,1A\\\hline
6,1A,C& 6,1D,1&6,1E,12&7,0,1A&7,1,C&7,4,1&
7,6,4&7,8,7&7,F,2&7,13,8& 7,14,19&7,18,0&8,1,3\\\hline
8,2,14&8,3,6& 8,4,17&8,8,1D&8,A,1&8,D,15&
8,F,18&8,14,10&8,15,2&8,17,5&8,19,8& 8,1A,19&8,1E,0\\\hline
8,1F,11&9,0,8&9,1,19& 9,4,E&9,8,14&9,C,1A&9,11,12&
9,12,4&9,14,7&9,15,18&9,16,A&9,1C,13& A,1,10\\\hline
A,3,13&A,4,5&A,7,19& A,9,1C&A,C,11&A,D,3&A,12,1A&A,13,C&
A,18,4&A,1B,18&B,0,15&B,3,A&B,4,1B\\\hline
B,7,10&B,8,2&B,10,E&B,11,0& B,18,1A&B,1C,1&C,1,1D&C,D,10&C,13,19&
C,14,B&C,1D,9&D,9,1&D,10,1B\\\hline

D,13,10&E,4,0&E,6,3&E,8,6&E,A,9& E,C,C&E,F,1&E,10,12&E,11,4&
E,14,18&E,19,10&E,1E,8&F,1,2\\\hline F,A,0&F,C,3&
10,0,7&10,3,1B&10,A,16&10,B,8&10,C,19&10,D,B&10,10,0&10,13,14&10,14,6&
11,3,12&11,4,4\\\hline 11,5,15&11,6,7&11,8,A&
11,D,2&11,E,13&11,10,16&11,13,B&
11,16,0&11,17,11&11,19,14&11,1A,6&12,2,17& 12,5,C\\\hline
12,C,7&12,D,18&12,E,A& 12,10,D&12,12,10&12,13,2&12,14,13&12,18,19&
12,19,B&12,1C,0&12,1E,3&13,0,B&13,A,1A\\\hline

13,E,1&13,13,18&13,1D,8&14,2,5&
14,4,8&14,9,0&14,C,14&14,11,C&14,16,4&
14,18,7&14,19,18&14,1E,10&15,0,18\\\hline

15,2,1B&15,8,5&15,C,B&15,F,0&15,11,3&
15,15,9&16,0,F&16,1,1&16,4,15&16,8,1B&16,C,2&16,11,19&16,15,0\\\hline
16,16,11&16,18,14&17,2,9&17,7,1&17,8,12&17,9,4&
17,C,18&17,11,10&17,16,8&18,1,E& 18,5,14&18,6,6&18,8,9\\\hline
18,9,1A&18,D,1& 18,11,7&18,13,A&18,17,10&18,18,2&
18,19,13&19,0,13&19,1,5&19,3,8&19,4,19& 19,5,B&19,C,6\\\hline
19,10,C&19,13,1&19,14,12& 19,15,4&19,19,A&19,1D,10&19,1E,2&
1A,0,A&1A,1,1B&1A,6,13&1A,8,16&1A,E,0& 1A,10,3\\\hline
1A,12,6&1A,19,1&1A,1A,12& 1B,0,1&1B,1,12&1B,2,4&1B,5,18&1B,6,A&
1B,A,10&1B,B,2&1C,0,17&1C,2,1A&1C,8,4\\\hline
1C,C,A&1C,10,10&1C,12,13&1C,15,8&1C,1A,0&
1C,1C,3&1D,0,E&1D,2,11&1D,3,3&1D,11,18&
1D,12,A&1E,2,8&1E,7,0\\\hline
1E,8,11&1E,9,3&1E,14,4&1E,18,A&1E,1C,10&1F,4,2&
1F,8,8&1F,D,0&&&&&\\\hline
\end{tabular}
\end{center}
\end{table*}

\begin{table*}
\renewcommand{\arraystretch}{1.2}
\begin{center}
\caption{Decoding Procedures While Applying IFBMA} \label{tableII}
\begin{tabular}{|c|c|c|c|c|c|} \hline
$k$&$\Delta^{(k)}$&$C^{(k)}(x)$&$A^{(k)}(x)$&$l^{(k)}$&$\gamma^{(k)}$\\
\hline 0&n.a.&1&1&0&1\\ \hline 1&$\alpha^4$&$1+\alpha^4x$
&1&1&$\alpha^4$\\
\hline 2&0&$\alpha^4+\alpha^8x$
&$x$&1&$\alpha^4$\\
\hline 3&$\alpha^9$&$\alpha^8+\alpha^{12}x+\alpha^{9}x^2$
&$\alpha^4+\alpha^8x$&2&$\alpha^9$\\
\hline 4&0&$\alpha^{17}+\alpha^{21}x+\alpha^{18}x^2$
&$\alpha^4x+\alpha^8x^2$&2&$\alpha^9$\\
\hline
5&$\alpha^{17}$&$\alpha^{26}+\alpha^{30}x+\alpha^{17}x^2+\alpha^{25}x^3$
&$\alpha^{17}+\alpha^{21}x+\alpha^{18}x^2$&3&$\alpha^{17}$\\
\hline 6&0&$\alpha^{12}+\alpha^{16}x+\alpha^{3}x^2+\alpha^{11}x^3$
&$\alpha^{17}x+\alpha^{21}x^2+\alpha^{18}x^3$&3&$\alpha^{17}$\\
\hline
\end{tabular}
\end{center}
\end{table*}

\begin{table*}
\renewcommand{\arraystretch}{1.2}
\begin{center}
\caption{List 79 $(i,j,k,u)$ elements of $A_2$} \label{tableIII}
\begin{tabular}{|c|c|c|c|c|c|c|c|c|c|c|c|c|} \hline
0,2,A,3 & 0,4,6,5 & 0,5,4,6 & 0,6,2,7 & 0,8,D,9 & 0,9,B,A &
0,A,9,B & 0,B,7,C & 0,C,5,D & 0,D,3,E & 0,E,1,0 & 0,E,1,F &
1,1,9,4 \\\hline

1,5,1,8 & 1,6,E,9 & 1,D,0,1 & 1,E,D,2 & 2,3,2,8 & 2,5,D,A &
2,C,E,2 & 2,D,C,3 & 2,E,A,4 & 3,1,3,8 & 3,4,C,B & 3,8,4,0 &
3,9,2,1 \\\hline 3,C,B,4 & 4,3,B,C & 4,6,5,0 & 4,B,A,5 & 4,C,8,6 &
4,E,4,8 & 5,2,A,D & 5,5,4,1 & 5,8,D,4 & 5,A,9,6 & 5,C,5,8 &
6,1,9,E & 6,2,7,0 \\\hline 6,8,A,6 & 6,9,8,7 & 6,A,6,8 & 7,0,8,F &
7,1,6,1 & 7,8,7,8 & 8,6,8,8 & 8,7,6,9 & 8,8,4,A & 8,A,0,C &
8,E,7,1 & 9,6,5,A & 9,8,1,C \\\hline 9,B,A,0 & 9,C,8,1 & 9,D,6,2 &
A,2,A,8 & A,4,6,A & A,5,4,B & A,6,2,C & A,C,5,3 & B,0,B,8 &
B,4,3,C & B,7,C,0 & B,8,A,1 & B,B,4,4 \\\hline C,3,2,D & C,4,0,E &
C,A,3,5 & D,0,5,C & D,2,1,E & D,3,E,0 & D,4,C,1 & D,9,2,6 &
E,0,2,E & E,1,0,F & E,8,1,7 & F,0,E,1 & F,7,0,8 \\\hline F,E,1,0 &
& &&&&&&&&&&\\\hline

\end{tabular}
\end{center}
\end{table*}

\begin{table*}
\renewcommand{\arraystretch}{1.2}
\begin{center}
\caption{List 190 $(i,j,k,u)$ elements of $A_3$} \label{tableIV}
\begin{tabular}{|c|c|c|c|c|c|c|c|c|c|c|c|c|} \hline
0,1,B,3 & 0,2,9,4 & 0,3,7,5 & 0,6,1,8 & 0,7,E,9 & 0,D,2,0 &
1,1,8,5 & 1,3,4,7 & 1,5,F,9 & 1,6,D,A & 1,7,B,B & 1,8,9,C &
1,9,7,D \\
\hline 1,A,5,E & 1,B,3,F  & 1,F,A,4 & 2,3,1,9 & 2,6,A,C & 2,7,8,D
& 2,8,6,E & 2,9,4,F & 2,A,2,1 & 2,B,F,2 & 2,F,7,6 &
3,1,2,9 & 3,2,F,A \\
\hline 3,3,D,B & 3,4,B,C & 3,5,9,D & 3,6,7,E & 3,7,5,0 & 3,7,5,F &
3,8,3,1 & 3,A,E,3 & 3,B,C,4 & 3,C,A,5 &
3,E,6,7 & 3,F,4,8 & 4,1,E,B \\
\hline 4,4,8,E & 4,5,6,0 & 4,5,6,F & 4,6,4,1 & 4,7,2,2 & 4,9,D,4 &
4,B,9,6 & 4,C,7,7 & 4,F,1,A
& 5,1,B,D & 5,2,9,E & 5,3,7,F & 5,4,5,1 \\
\hline 5,6,1,3 & 5,7,E,4 & 5,9,A,6 & 5,C,4,9 & 5,E,F,B & 5,F,D,C &
6,0,A,E & 6,1,8,0
&  6,3,4,2 & 6,5,F,4 & 6,6,D,5 & 6,7,B,6 & 6,8,9,7 \\
\hline 6,A,5,9 & 6,B,3,A & 6,D,E,C & 6,E,C,D & 6,F,A,E & 7,1,5,2 &
7,2,3,3
& 7,4,E,5 & 7,5,C,6 & 7,6,A,7  & 7,8,6,9 & 7,9,4,A & 7,A,2,B \\
\hline

7,B,0,C & 7,B,F,C & 7,C,D,D & 7,D,B,E & 7,E,9,F & 7,F,7,1 &
8,1,2,4 & 8,3,D,6 & 8,4,B,7 & 8,7,5,A & 8,A,E,D & 8,B,C,E
& 8,C,A,F \\
\hline 8,F,4,3 & 9,2,C,7 & 9,3,A,8 & 9,5,6,A & 9,7,2,C & 9,8,F,D &
9,9,D,E & 9,A,B,0 & 9,A,B,F & 9,B,9,1 & 9,D,5,3
& 9,E,3,4 & 9,F,1,5 \\
\hline A,0,D,7  & A,3,7,A & A,4,5,B & A,7,E,E & A,8,C,F & A,A,8,2
& A,B,6,3 & A,D,2,5 & A,E,F,6 & A,F,D,7
& B,2,6,B & B,4,2,D & B,5,0,E \\
\hline B,5,F,E & B,6,D,0 & B,6,D,F  &
B,7,B,1 & B,8,9,2 & B,9,7,3 & B,B,3,5 & B,C,1,6 & B,D,E,7 & B,E,C,8 & B,F,A,9 & C,0,7,B & C,1,5,C \\
\hline C,2,3,D & C,3,1,E & C,4,E,F & C,6,A,2 & C,7,8,3  & C,8,6,4
& C,9,4,5 & C,B,F,7 &
C,C,D,8 & C,D,B,9 & C,E,9,A & C,F,7,B & D,0,4,D \\
\hline D,2,0,F & D,2,F,0 & D,2,F,F & D,3,D,1 & D,4,B,2 & D,5,9,3 &
D,6,7,4  & D,7,5,5 & D,9,1,7 & D,A,E,8 & D,B,C,9 & D,C,A,A &
D,D,8,B
 \\
\hline D,E,6,C & D,F,4,D & E,0,1,F & E,2,C,2 & E,3,A,3 & E,4,8,4 &
E,5,6,5 & E,6,4,6 & E,7,2,7  &  E,8,0,8 & E,8,F,8 & E,9,D,9 & E,A,B,A \\
\hline E,B,9,B & E,C,7,C & E,D,5,D & E,E,3,E & E,F,1,F & F,1,B,3 &
F,2,9,4 & F,3,7,5 & F,4,5,6 & F,5,3,7 & F,7,E,9  &
F,8,C,A & F,9,A,B \\
\hline F,A,8,C & F,B,6,D & F,C,4,E & F,D,2,0 & F,D,2,F & F,E,0,1 &
F,E,F,1 & F,F,D,2 & &&&& \\ \hline

\end{tabular}
\end{center}
\end{table*}

\section*{Acknowledgment}
The authors wish to thank Dr. Jian-Hong Chen for his assistance in
providing examples and computing services of MVIF.

\end{document}